\documentstyle[12pt,psfig]{article}
\def\be{\begin{equation}}
\def\ee{\end{equation}}
\def\tr{{\rm Tr}}

\begin{document}

\title{\large NEGATIVE ENTROPY IN QUANTUM INFORMATION THEORY\thanks{To
appear in the Proceedings of the 2nd International Symposium on
Fundamental Problems in Quantum Physics, Oviedo 1996, Kluwer Academic
Publishers.}}
\author{Nicolas J. Cerf and Chris Adami\\[3mm]
\it Kellogg Radiation Laboratory, 106-38\\
\it California Institute of Technology\\
\it Pasadena, California 91125}
\date {}
\maketitle

\noindent 
We present a quantum information theory that
allows for the consistent description of quantum entanglement.
It parallels classical (Shannon) information theory but 
is based entirely on density matrices, rather than probability 
distributions, for the description of quantum ensembles.
We find that, unlike in Shannon theory,
conditional entropies can be negative when considering
quantum entangled systems such as an Einstein-Podolsky-Rosen pair,
which leads to a violation of well-known bounds of classical
information theory. Negative quantum entropy can be traced back
to ``conditional'' density matrices which admit eigenvalues larger
than unity. A straightforward definition of mutual quantum entropy,
or ``mutual entanglement'', can also be constructed using a ``mutual''
density matrix. Such a unified information-theoretic description of
classical correlation and quantum entanglement
clarifies the link between them: the latter can be
viewed as ``super-correlation'' which can induce classical
correlation when considering a ternary or larger system.\\[7mm]

\noindent{\bf 1. INTRODUCTION}\bigskip

Quantum information theory~\cite{bib_bennett} is a new field with
potential implications for the conceptual foundations
of quantum mechanics. It appears to be the basis for a proper understanding
of the emerging fields of quantum computation~\cite{bib_divincenzo},
quantum communication~\cite{bib_superdens}, and quantum 
cryptography~\cite{bib_crypto}.
Although some fundamental results have been obtained recently
such as the quantum noiseless coding theorem~\cite{bib_schum}
or the rules governing the extraction of classical information
from quantum entropy, it is still puzzling in many respects.
Quantum information processing basically 
deals with quantum bits (qubits)~\cite{bib_schum} rather than bits,
the former obeying quantum laws quite
different from the classical physics of bits that we are used to. Most 
importantly, qubits can exist in quantum {\em superpositions}, a notion 
completely inaccessible to classical mechanics, or even classical thinking. 
To accommodate the relative phases in quantum superpositions, quantum 
information theory must be based on mathematical constructions which reflect
these: the density matrices. The central object of 
information theory, the entropy, has been introduced in 
quantum mechanics by von Neumann~\cite{vonneumann} 
\be
S(\rho) = -\tr\,\rho\log\rho\;. \label{vnentropy}
\ee
Its relationship to the Boltzmann-Gibbs-Shannon entropy
\be
H({\rm\vec{p}}) = - \sum_i p_i\log p_i \label{bgsentropy}
\ee
is obvious when considering the von Neumann entropy of a mixture
of orthogonal states, in which
case the density matrix $\rho$ in (\ref{vnentropy}) contains classical 
probabilities $p_i$ on its diagonal, and 
$S(\rho)=H({\rm\vec{p}})$. In 
general, however, quantum mechanical density matrices have off-diagonal terms,
which reflect the relative quantum phases in superpositions. 
\par

In classical (Shannon) information theory~\cite{bib_shannon}
the concept of conditional
probabilities has given rise to the definition of conditional and mutual
entropies. These can be used to elegantly describe the trade-off between 
entropy and information in measurement, as well as the characteristics of a 
transmission channel. For example, for two systems $A$ and $B$, 
the measurement of $A$ by $B$ is expressed by the equation for the entropies
\be
H(A) = H(A|B) + H(A{\rm:}B)\;. \label{conserv}  
\ee
Here, $H(A|B)$ is the entropy of $A$ after having measured those pieces
that become correlated in $B$, while
$H(A{\rm:}B)$ is the information gained about $A$ via the measurement of $B$.
Mathematically, $H(A|B)$ is a {\em conditional} entropy,
and is defined using the conditional probability $p_{i|j}$ and the 
joint probability $p_{ij}$ describing random variables from ensembles $A$
and $B$:
\be    \label{classcond}
H(A|B) = -\sum_{ij}p_{ij}\log p_{i|j}\;.
\ee
The mutual entropy or information $H(A{\rm:}B)$, 
on the other hand,
is defined via the mutual probability $p_{i:j} = p_i\,p_j/p_{ij}$ as 
\be
H(A{\rm:}B) = -\sum_{ij} p_{ij} \log p_{i:j}\;. 
\ee
Simple relations such as $p_{ij} = p_{i|j}\,p_j$ imply equations such as
$H(A|B)=H(AB)-H(B)$ and all the other usual relations of classical information
theory [{\it e.g.}, Eq. (\ref{conserv})].
Curiously, a quantum information theory paralleling these
constructions has never been attempted. Rather, a ``hybrid'' theory
was used in which quantum probabilities
are inserted in the classical formulae given above, thereby loosing
the quantum phase crucial to density matrices
(see, e.g.,~\cite{bib_zurek}). 
Below in Section~2 we show that a consistent quantum information theory can be
developed that parallels the construction outlined above, while based
entirely on matrices~\cite{bib_caneginfo}.\\[7mm]

\noindent{\bf 2. QUANTUM INFORMATION THEORY}\bigskip

Let us consider the information-theoretic description of a
composite quantum system $AB$. A straightforward quantum generalization
of Eq.~(\ref{classcond}) suggests the definition
\be  \label{eq_defcond}
S(A|B)= - \tr_{AB} [ \rho_{AB} \log \rho_{A|B} ]
\ee
for the quantum conditional entropy. In order for
such an expression to hold, we define the concept of
a ``conditional'' density matrix,
\be  \label{eq_condmat}
\rho_{A|B} = \left[ \rho_{AB}^{1/n} ({\bf 1}_A \otimes \rho_B)^{-1/n}
             \right]^n  \qquad n\to \infty  \;,
\ee
the analog of the conditional probability $p_{i|j}$. Here, 
${\bf 1}_A$ is the unity matrix in the Hilbert space for $A$,  $\otimes$
stands for the tensor product in the joint Hilbert space, and
$\rho_B= \tr_A [\rho_{AB}]$
denotes a ``marginal'' density matrix, analogous to the marginal
probability $p_j=\sum_i p_{ij}$. The peculiar form involving the infinite
limit in Eq.~(\ref{eq_condmat})
is necessary because joint and marginal density matrices
do not commute in general. However, 
the definition implies that the standard relation
\be
S(A|B)=S(AB)-S(B)
\ee
holds for the quantum entropies and that $S(A|B)$ is invariant under
any unitary transformation of the product form $U_A \otimes U_B$.
More precisely, it is easy to see that $\rho_{A|B}$ is
a positive Hermitian operator (in the joint Hilbert space)
whose spectrum is invariant under $U_A\otimes U_B$.
Despite the apparent similarity between
the quantum definition for $S(A|B)$ and the standard classical one
for $H(A|B)$, dealing with matrices rather than scalars
opens up a quantum realm for information theory exceeding
the classical one.
The crucial point is that, while $p_{i|j}$ is a probability distribution
in $i$ (in particular $0\le p_{i|j} \le 1$), its quantum analog $\rho_{A|B}$ 
is {\em not} a density operator: it can have eigenvalues {\em larger} 
than one,
and, consequently, the associated conditional entropy $S(A|B)$ can
be {\em negative}. Only such a matrix-based quantum formalism
consistently accounts for the well-known non-monotonicity of quantum
entropies~(see, {\it e.g.}, \cite{bib_wehrl}).
This means that it is acceptable, in quantum information
theory, to  have $S(AB) < S(B)$, {\it i.e.},
the entropy of the entire system $AB$
can be smaller than the entropy of one of its subparts $B$,
a situation which is of course forbidden in classical information theory.
This happens for example in the case of quantum {\em entanglement} between 
$A$ and $B$, and will be illustrated below for an EPR pair. Note that, as
a consequence of the concavity of $S(A|B)$, a property related to strong
subadditivity (see, {\it e.g.}, \cite{bib_wehrl})
any separable state ({\it i.e.}, a mixture of product states)
is associated with non-negative $S(A|B)$. (The converse is not true.) 
Therefore, the non-negativity of conditional entropies can be viewed as a
necessary condition for separability, and we have shown that this
condition can be related to entropic Bell inequalities~\cite{bib_cabell}.
\par

Similarly, the quantum analog of the mutual entropy can be
constructed, defining a ``mutual'' density matrix
\be   \label{eq_mutmat}
\rho_{A:B} = \left[  (\rho_A \otimes \rho_B)^{1/n}\rho_{AB}^{-1/n}
             \right]^n  \qquad n\to \infty  \;,
\ee
the analog of the mutual probability $p_{i:j}$. As previously, this
definition implies the standard relation 
\be
S(A{\rm:}B)=S(A)+S(B)-S(AB)
\ee
between the quantum entropies. This definition extends
the classical notion of mutual or {\em correlation}
entropy $H(A{\rm:}B)$ to the quantum notion
of mutual {\em entanglement} $S(A{\rm:}B)$ and applies 
to pure as well as mixed states; $S(A{\rm:}B)$ is a general measure
of correlations {\em and} ``super-correlations'' in information theory.
In fact, all the above quantum
definitions reduce to the classical ones for a diagonal $\rho_{AB}$, which
suggests that Eqs. (\ref{eq_condmat}) and (\ref{eq_mutmat})
are very reasonable assumptions. It is possible that
other definitions of $\rho_{A|B}$
and $\rho_{A:B}$ could be proposed, but we believe this choice is simplest. 
This formalism suggests that all the relations between classical entropies
({\it e.g.}, the chain rules for entropies and mutual entropies)
also have a quantum analog, and we make
use of it in~\cite{bib_cabell,bib_cameasure}.
\par

\begin{figure}
\centerline{\psfig{file=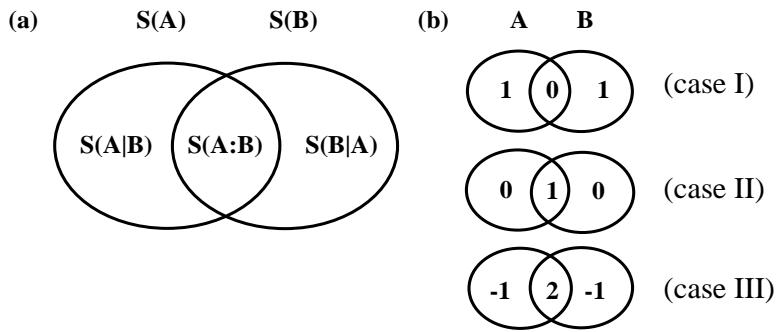,width=4.00in,angle=-90}}
\caption {(a) General entropy diagram for a quantum composite system $AB$.
(b) Entropy diagrams for three cases of a system of 2 qubits: (I) independent,
(II) classically correlated, (III) quantum entangled. \label{figAB} }
\end{figure}

The relations between entropies are conveniently summarized
by a Venn-like entropy diagram, as shown in Fig.~\ref{figAB}a.
The important difference
between classical and quantum entropy diagrams is that the basic
inequalities relating the entropies are ``weaker'' in the quantum case,
allowing for negative conditional entropies and ``excessive'' mutual 
entropies~\cite{bib_caneginfo}.
For example, the upper bound for the mutual entropy (which is directly
related to the channel capacity) is 
$H(A{\rm:}B) \le \min[H(A),H(B)]$
in classical information theory, while it can reach twice the classical
upper bound $S(A{\rm:}B) \le  2 \min[S(A),S(B)]$
in quantum information theory as a consequence of the Araki-Lieb inequality
(see, {\it e.g.}, \cite{bib_wehrl}).
In Fig.~\ref{figAB}b, we show the entropy diagram corresponding to three
limiting cases of a composite system of two dichotomic variables
(e.g., 2 qubits): independent variables (case I),
classically correlated variables (case II), and quantum entangled
variables (case III). In all three cases, each subsystem taken separately
is in a mixed state of entropy $S(A)=S(B)=1~$bit.
Cases I and II correspond to classical situations
(which can of course be described in our formalism with density
matrices as well), while case III is a purely quantum situation which violates
the bounds of classical information theory~\cite{bib_caneginfo}. 
It corresponds to an EPR pair, characterized by the pure state
$| \psi_{AB} \rangle = 2^{-1/2} (|01\rangle - |10\rangle)$,
and, accordingly, it is associated with a vanishing combined entropy
$S(AB)=0$. Using $\rho_{AB}=|\psi_{AB}\rangle \langle \psi_{AB}|$,
we see that subpart $A$ (or $B$)
has the marginal density matrix
$ \rho_A = \tr_B [\rho_{AB}] =
       {1\over 2} (|0\rangle \langle0| +|1\rangle \langle1|)$,
and is therefore in a mixed state of positive entropy. This purely
quantum situation corresponds to the unusual entropy diagram (--1,2,--1)
shown in Fig.~\ref{figAB}b. That the EPR situation cannot be described
classically is immediately apparent when considering the 
conditional density matrix\footnote{Note that for
Bell states, joint and marginal density matrices commute, simplifying
definitions (\ref{eq_condmat}) and (\ref{eq_mutmat}).}: indeed the latter
can be written as
\be   \label{eq_condmatex}
\rho_{A|B}= \rho_{AB} ({\bf 1}_A \otimes \rho_B)^{-1} = \left(
\begin{array}{rrrr}
0 & 0 & 0 & 0 \\
0 & 1 & -1 & 0 \\ 
0 & -1 & 1 & 0 \\ 
0 & 0 & 0 & 0
\end{array}  \right) \;.
\ee 
Plugging (\ref{eq_condmatex}) into
definition (\ref{eq_defcond}) immediately yields
$S(A|B)=-1$. This is a direct consequence of the fact
that $\rho_{A|B}$ has one ``unclassical'' ($>1$) eigenvalue, 2.
It is thus misleading to describe an EPR pair (or any of the Bell states)
as a correlated state within Shannon theory, since negative
conditional entropies are crucial to its description. In~\cite{bib_caneginfo},
we suggest that EPR pairs are better understood in terms of a qubit-antiqubit
pair, where the qubit (antiqubit) carries plus (minus) one bit of
information, and antiqubits are interpreted as qubits traveling 
{\it backwards} in time. Still, classical {\em correlations} (case II)
emerge when {\em observing} an entangled EPR pair. Indeed, after measuring 
$A$, the outcome of the measurement of $B$ is known with 100\% certainty.
The key to this discrepancy lies in the information-theoretic
description of the measurement process~\cite{bib_cameasure}
and will be briefly addressed in the next section.\\[7mm]

\noindent{\bf 3. CORRELATION VERSUS ENTANGLEMENT}\bigskip

\begin{figure}
\centerline{\psfig{file=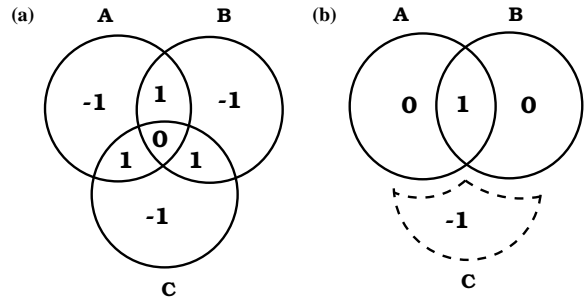,width=3.00in,angle=-90}}
\caption{(a)~Ternary entropy diagram for an ``EPR-triplet''.
(b)~Entropy diagram for subsystem $AB$ unconditional on $C$. \label{figABC} }
\end{figure}

The concept of negative conditional entropy turns out to be very
useful to describe $n$-body composite quantum systems, and it
sheds new light on the creation of classical correlations from quantum
entanglement. Consider for example a 3-body system $ABC$
in a GHZ state (or an ``EPR-triplet''),
$| \psi_{ABC} \rangle = 2^{-1/2} (|000\rangle + |111\rangle)$.
As it is a pure (entangled) state, the combined entropy is $S(ABC)=0$. 
The corresponding ternary entropy diagram of $ABC$ is shown
in Fig.~\ref{figABC}a. Note that the {\it ternary} mutual entropy
$S(A{\rm:}B{\rm:}C)=S(A)+S(B)+S(C)-S(AB)-S(AC)-S(BC)+S(ABC)$ vanishes
(see the center of the diagram); this
is generic to any fully entangled three-body system.
When tracing over the degree of freedom associated with $C$, say, the
resulting marginal density matrix for subsystem $AB$ is
$\rho_{AB} = \tr_C [\rho_{ABC}] =
       {1\over 2} (|00\rangle \langle 00| +|11\rangle \langle 11|)$,
corresponding to a classically correlated system (case II). As the
density matrix {\em fully} characterizes a quantum system, subsystem $AB$
(unconditional on $C$, i.e., ignoring the existence of $C$)
is in this case physically {\em indistinguishable} from a statistical ensemble
prepared with an equal number of $|00\rangle$ and $|11\rangle$ states.
Thus, $A$ and $B$ are correlated in the sense of Shannon theory
if $C$ is ignored. 
The ``tracing over'' operation depicted in Fig.~\ref{figABC}b
illustrates this creation of
classical correlation from quantum entanglement. This feature
is central to description the measurement process that we propose
in \cite{bib_cameasure}, where $A$ and $B$ represent two parts of the
measurement device, while $C$ is the measured quantum system.
The subsystem $AB$ unconditional on $C$
has a positive entropy $S(AB)=1$~bit, and is indistinguishable
from a  classical correlated mixture (this corresponds to the
generation of random numbers). On the other hand, the entropy 
of $C$ conditional on $AB$, $S(C|AB)$, is negative and equal to $-1$~bit,
thereby counterbalancing $S(AB)$ to yield a vanishing combined entropy
\be
S(ABC)=S(AB)+S(C|AB)=0  \;.
\ee 
as expected in view of the quantum entanglement between $AB$ and $C$.
We suggest in \cite{bib_cameasure} that
this information-theoretic interpretation of entanglement
paves the way to a natural, unitary, and causal model of the measurement
process, devoid of any assumption of a wave-function collapse,
while implying all the well-known results of conventional 
probabilistic quantum mechanics. 
The same framework can also be used to interpret the observation of classical
correlation between the measurement devices that occurs
in the measurement of an EPR pair, and sheds 
new light on quantum paradoxes~\cite{bib_acparadoxes}.
\par
\medskip
We would like to thank H. Bethe and A. Peres for very useful discussions.
This work was supported in part by the National Science Foundation  
Grant Nos. PHY91-15574 and PHY94-12818.\\


\begin{thebibliography}{99}
\baselineskip 13pt
\bibitem{bib_bennett} C. H. Bennett, {\it Physics Today}
 {\bf 48}(10), 24 (1995).
\bibitem{bib_divincenzo} C. H. Bennett and D. P. DiVincenzo, {\it Nature}
 {\bf 377}, 389 (1995); D. P. DiVincenzo, {\it Science} {\bf 270}, 255 (1995).
\bibitem{bib_superdens} C. H. Bennett and  S. J. Wiesner, {\it Phys. Rev.
  Lett.} {\bf 69}, 2881 (1992); C. H. Bennett {\it et al.}, {\it Phys. Rev.
  Lett.} {\bf 70}, 1895 (1993).
\bibitem{bib_crypto} A. Ekert, {\it Nature} {\bf 358}, 14 (1992);
C. H. Bennett, G. Brassard, and N. D. Mermin, {\it Phys. Rev. Lett.} {\bf 68},
557 (1992).
\bibitem{bib_schum} B. Schumacher, {\it Phys. Rev. A} {\bf 51}, 2738 (1995);
R. Jozsa and B. Schumacher, {\it J. Mod. Opt.} {\bf 41}, 2343 (1994).
\bibitem{vonneumann} J. von Neumann, {\em Mathematische Grundlagen 
der Quantenmecha\-nik}, Springer Verlag, Berlin (1932).
\bibitem{bib_shannon} C. Shannon and W. Weaver,
{\em The Mathematical Theory of Communication}, University of Illinois Press, 
Urbana (1949).
\bibitem{bib_zurek} W. H. Zurek (ed.), {\em Complexity, Entropy 
and the Physics of Information}, Santa Fe Institute Studies in the Sciences
of Complexity Vol. VIII, Addison-Wesley (1990).
\bibitem{bib_caneginfo} N. J. Cerf and C. Adami, ``Negative 
entropy and information in quantum mechanics'', e-print quant-ph/9512022;
``Quantum information theory of entanglement'', e-print quant-ph/9605039.
\bibitem{bib_wehrl} A. Wehrl, {\it Rev. Mod. Phys.} {\bf 50}, 221 (1978).
\bibitem{bib_cabell} N. J. Cerf and C. Adami, ``Entropic Bell inequalities'',
e-print quant-ph/9608047.
\bibitem{bib_cameasure} N. J. Cerf and C. Adami, ``Quantum  
mechanics of measurement'', e-print quant-ph/9605002.
\bibitem{bib_acparadoxes} C. Adami and N. J. Cerf, Caltech preprint
KRL--MAP--204 (1996).
\end{thebibliography}
\end{document}